\documentclass[twocolumn,showpacs,amsmath,amssymb]{revtex4}
\usepackage{graphicx}
\usepackage{dcolumn}
\usepackage{bm}

\newcommand{\ba}{\begin{eqnarray}}
\newcommand{\ea}{\end{eqnarray}}
\newcommand{\be}{\begin{equation}}
\newcommand{\ee}{\end{equation}}
\newcommand{\bd}{\begin{displaymath}}
\newcommand{\ed}{\end{displaymath}}

\begin{document}

\draft

\title{Observable form of pulses emitted from relativistic
collapsing objects}
\author{Valeri P. Frolov$^*$ and and Hyun Kyu Lee$^{*,\dagger, \S}$}
\address{
  \medskip
  $^*$Theoretical Physics Institute,Department of Physics, University of
  Alberta,  Edmonton, AB, Canada, T6G 2J1\\
  {\rm E-mail: \texttt{frolov@phys.ualberta.ca}} \\
  $^\dagger$ KIPAC,  2575 Sand Hill Rd, Menlo Park, CA 94025, USA \\ $^{\S}$ Department of Physics, Hanyang University,Seoul
  133-791, Korea\\
  {\rm E-mail: \texttt{hyunkyu@hanyang.ac.kr}}
  \medskip
}

\date{\today}

\begin{abstract}
In this work, we  discuss observable characteristics of the
radiation emitted from a  surface of a collapsing object. We study
a simplified model in which a radiation of massless particles has
a sharp in time profile and it happens at the surface at the same
moment of comoving time.  Since the radiating surface has finite
size the observed radiation will occur during some finite time.
Its redshift and bending angle are affected by the
strong gravitational field. We obtain a simple expression for the
observed flux of the radiation as a function of time. To find an
explicit expression for the flux we develop an analytical
approximation for the bending angle and time delay for null rays
emitted by a collapsing surface. In the case of the bending angle
this approximation is an improved version of the earlier proposed
Beloborodov-Leahy-approximation. For rays emitted at $R > 2R_g$
the accuracy of the proposed improved approximations for the
bending angle and time delay is of order (or less) than 2-3$\%$.
By using this approximation  we obtain an approximate analytical
expression for the observed flux and study its properties.
\end{abstract}

\pacs{04.40.Dg, 95.85.-e, 97.60.-s \hfill
Alberta-Thy-19-04/SLAC-PUB-10998}

\maketitle

\section{Introduction}

Neutron stars and black holes are compact relativistic objects.
Light propagating in the vicinity of such objects is affected by a
strong gravitational field. For the description of the photon
propagation under these conditions the general relativity is
required. This is evident for black holes where the gravitational
potential  at the horizon (in the $c^2$ units) is of order of 1.
Though a value of the gravitational potential at a surface
($R_{NS} \sim 8M_{\odot}$) of neutron stars ($M_{NS} \sim 1.5
M_{\odot}$) \cite{lp1}  is smaller and is of  order of $\sim 0.2$,
it was demonstrated recently  that general relativistic effects
might be important for understanding the features of the radiation
coming from these objects \cite{pfc}-\cite{bel}.  In particular,
one of the explanations of X-ray bursts  is connected with
possible explosions  of the matter which is
accumulated on the surface of a neutron star  entering in a binary
system as a result of ``sucking" the matter from a usual-star
companion. It was shown  (see \cite{ll}, \cite{mr}, \cite{clm})
that the profile of the X-ray-curve observed by a distant observer
can be explained only if the effects of general relativity are
taken into account. In particular, according to the general
relativity, because of the gravitational bending of light rays, a
distant observer can see a part of the opposite side of the
neutron star which is invisible in a flat spacetime.   The
radiation emitted from this part gives an important contribution
and has a visible impact on the form of the signal from the X-ray
burst.

The effects of the general relativity also modify considerably
light curves for continuous in time radiation from the surface of
a collapsing star as seen by a distant observer. These effects
were studied in detail by Ames and Thorne \cite{at},
Jaffe \cite{jaffe} and Lake \cite{lr}.  In this study main
attention was focused on the details connected with light emitted
near  (unstable) circular photon orbits at $3r_g/2$, where $r_g$
is a gravitational radius of a collapsing star. In  such
considerations there were usually adopted a number of simplifying
assumptions, such as:  (1) Spherical geometry; (2) Dust-like
(pressure free) equation of state; (3) Radiation comes only from
the (free-falling) surface of the star; and (4) It is continuous
in time.

In this paper we would like to discuss a slightly different set
up, when the assumption (4) is violated. Namely we assume that the
radiation emitted from the surface of a collapsing spherically
symmetric stellar object  has a profile of a sharp  in time pulse.
Such radiation may occur in different situations. For example,
suppose a neutron star or a proto-neutron star looses its
stability as a result of the accretion of matter onto it or due to
the softening of   equation of state \cite{kaon} at  the center
which is supposed to be already several times higher than the
nuclear density .  During the collapse, the matter density of a
compact object  is growing and  the whole system   evolves into
the much higher density  than the normal nuclear density
\cite{sha}--\cite{nsc}, beyond which new hadronic phase transitions
might take place \cite{schafer}. One might expect   a
possible sharp-in-time emission of massless particles (photons and
neutrino) during such phase transition \cite{vro}.

Since at the moment there are a lot of theoretical uncertainties
in the estimations and detailed description of these processes, we
shall study a simplified (toy) model. Namely we assume that a
radiation of massless particles has a sharp in time pulse profile
and it happens at the surface at the same instant of time (from a
point of view of a comoving observer). The time required for the
radiation to reach a distant observer depends on the position of a
radiative region on the collapsing surface. For this reason the
pulse emission results in a continuous flux received by the
observer during some finite interval of time. During this interval
the flux as well as the redshift factor changes. In principle,
knowing the redshift and light curves allows one to obtain direct
information about the collapsing body at the moment when the
radiation occurs. We shall demonstrate that if the emission occurs
at the moment when the radius of a collapsing body is $R>4.5M$,
the observed characteristics of the pulse radiation with a very
high accuracy can be written in an explicit form in terms of
elementary functions. We shall use this result to study how
different (gravitational redshift, Doppler and focusing) effects
affect the form of redshift and light curves.

In section 2 we consider null rays emitted from a surface of a
spherical collapsing object. The equations of motion of the freely
collapsing surface and for rays propagating in the surrounding
Schwarzschild geometry are well known. In order to fix the notations
we collect the required formulas in section 2 in the form they are
used in the further sections. The time of arrival of null rays
emitted by a collapsing surface and their bending angle are given by
elliptic integrals. In section 3, we propose an explicit analytic
expressions which  approximate these quantities with very high
accuracy. For rays emitted at $R > 2R_g$ this accuracy is of order
(or less) than  2-3$\%$. This approximation is used to obtain the
expressions for the intensity and flux for a short flash observed  by
a distant observer (section~4). The results of numerical calculations
of the redshift, bending angle and light curves are given in section
5. In section 6, the results are summarized  and discussed.

\section{Basic Equations}

\subsection{Freely-falling  spherical surface}

We consider a photon emitted from a  collapsing spherical
surface and propagating to the observer at infinity in the
background of Schwarzschild metric
    \ba
    ds^2 = -f dt^2 + f^{-1} dr^2 + r^2 d\Omega
    \label{schw}\, ,
    \ea
\ba\label{ff} f=f(r) = 1- 2M/r\, , \ea where $M$ is the mass of
the collapsing object. We adopt the natural units, $c=G=\hbar=1$.
In \cite{at} and \cite{sha} the motion of a spherical surface
during the gravitational collapse was discussed under assumption
that the dynamical role of the pressure can be neglected. We also
use the dust-like equation of state so that a solution for the
interior of the collapsing object is a well known Tolman solution
\cite{tolmann}, while the surface follows a radial geodesic in the
Schwarzchild geometry \cite{os}.

Denote by $\tau$ the proper time as measured by an observer comoving
with the collapsing surface. We suppose that the collapse starts  at
$\tau=0$ and the initial surface radius is $R_0$.
Then, in the parametric form (using the conformal time parameter
$\eta$), the equation of motion of the surface is
    \ba
       R &=& \frac{R_0}{2}(1 + \cos \eta),\label{radius}\\
    \tau &=& \sqrt{\frac{R_0^3}{8M}}(\eta + \sin \eta)\label{tau}\, \,
    .
    \ea
In the same parametric form the Schwarzschild time $t^{(e)}$ corresponding
to the proper time $\tau$ at the surface is
\ba
t^{(e)} &=& 2M \ln \left|\frac{ \sqrt{R_0/2M -1} + \tan
(\eta/2)}{\sqrt{R_0/2M
    -1} - \tan (\eta/2)}\right| \nonumber \\
      && + 2M \sqrt{R_0/2M -1}[\eta + \frac{R_0}{4M}(\eta + \sin \eta)]
      \label{time}\, .
\ea
We assumed that $t^{(e)}(\tau=0)=0$.

We denote \ba v_i=f^{-1}(R)\frac{dR}{dt}\, . \ea This quantity has
the meaning of the invariant radial velocity. Then in a general
case  the  four-velocity of the collapsing  surface is \ba
    v^{\mu} = (dt^{(e)}/d\tau, dR/d\tau, 0, 0)\, ,
\ea where \ba {dt^{(e)}\over d\tau}={1\over
\sqrt{f(R)}\sqrt{1-v_i^2}}\, , \ea \ba\label{RRRR} {dR\over
d\tau}=v_i f(R){dt^{(e)}\over d\tau}={\sqrt{f(R)} v_i\over
\sqrt{1-v_i^2}}\, . \ea

Using eq.(\ref{radius}) and (\ref{time}),  we obtain the following
expression for the  velocity $v_i$ of a freely falling surface
with the initial radius $R_0$ at the moment when its radius is $R$
\ba v_i=-\sqrt{\frac{2M}{R}}
     \frac{\sqrt{1-R/R_0}}{\sqrt{1-2M/R_0}}\, . \label{v_s}
\ea We also have \ba\label{tetau} {dt^{(e)}\over
d\tau}={\sqrt{1-2M/R_0}\over 1-2M/R}\, , \ea \ba\label{Rtau}
{dR\over d\tau}=-\sqrt{2M/ R}\sqrt{1-R/R_0}\, . \ea

\subsection{Propagation of null rays emitted from the surface }

Consider a photon emitted from the surface. Its trajectory lies in
the plane. Without loss of generality we assume that it coincides
with a plane of the fixed coordinate $\phi$, so that the vector of the
4-momentum of the photon is
    \ba
    p^{\mu} &=& (p^t, p^r, p^{\theta}, 0)\, .
    \ea
Because of the symmetry of the Schwarzschild metric, $E=-p_t$ (the
energy at infinity) and $L=p_{\theta}$ (the angular momentum) are
constants of motion.  Instead of the angular momentum $L$ we shall
use  the impact parameter \ba l = L/E\, . \label{ldef} \ea The
radial momentum  $p^r$ is \ba p^r = \sigma E Z \, ,\label{pr} \ea
where \ba Z=Z(l,r)=\sqrt{1 - l^2 f(r)/r^2}\, . \ea Here and later
$\sigma$ denotes a sign function which takes the values $+$ and
$-$ for a forward ($p^r>0$) and backward ($p^r<0$) emission,
respectively.

There is an upper limit for the impact parameter, $l_{max}$, for a
photon which being  emitted  from a surface of radius $R$  reaches
the infinity:
    \ba
    l_{max} = \frac{R}{\sqrt{f(R)}}.
    \ea
Suppose a null ray emitted by a collapsing surface reaches the
infinity at $\theta=0$ and has the impact parameter $l$. Then  the
emission angle, $\beta$, as measured by an  observer comoving with
the surface, is defined by the relation
   \ba\label{beta}
   \cos \beta_{\sigma}(l,R) = \frac{ \sigma Z(l,R) - v_i}
           {1 -\sigma v_i Z(l,R)}\, .\label{beta}
   \ea
We consider only the light emitted with $\beta \leq \pi/2$.   A
tangentially emitted photon with respect to a comoving
observer($\beta = \pi/2$), for a collapsing surface ($v_i <0$), is
possible only for a backward emission. The corresponding impact
parameter $l=l_T$ is determined by the condition \ba\label{lT}
Z=-v_i\, . \ea Solving this equation one obtains
   \ba
     l_T   =\frac{R}{\sqrt{1- 2M/R_0}}\, . \label{fflt}
   \ea
For $R=3\sqrt{3}\sqrt{1- 2M/R_0}M $ the parameter $l_T$ is equal to
$3\sqrt{3}$, the value corresponding to the (unstable) circular
photon orbit. To escape delicacies connected with more complicated
behavior of the photons near this orbit we assume that $R >
3\sqrt{3}\sqrt{1- 2M/R_0}M$. In this case   for a photon which
reaches the infinity  the possible ranges of  an impact parameter are
$0 \leq l \leq l_{max}$ and $l_T \leq l \leq l_{max}$ for a forward
and backward emission, respectively. A discussion of the allowed
ranges of the impact parameter for the smaller radius up to $R \sim
2M$, can be found in  \cite{at} and \cite{jaffe}.

\subsection{Redshift}

Let $p_{\mu}^{(e)}$ be 4-momentum of a photon emitted by a
collapsing surface which has 4-velocity $v^{(e)\mu}$. Then
$\nu^{(e)}=-p_{\mu}^{(e)}v^{(e)\mu}$ is the frequency of the
photon as measured by a comoving to the surface observer. Since an
observer at infinity is at rest, its 4-velocity is
$v^{(o)\mu}=\delta^{\mu}_0$ and the observed frequency of the
photon is $\nu^{(o)}=-p_{\mu}^{(o)}v^{(o)\mu}$, where
$p_{\mu}^{(o)}$ is the 4-momentum of the photon at infinity. For a
given ray, the redshift factor $\Phi$ is defined as the ratio of
the emitted frequency $\nu^{(e)}$ to the observed at infinity
frequency $\nu^{(o)}$ \ba \Phi = {\nu^{(e)}\over \nu^{(o)}}\, .
\ea For a ray with the impact parameter $l$ emitted from the
surface of the radius $R$ one has \ba \Phi_{\sigma}(l,R) = {1
-\sigma v_i Z(l,R)\over \sqrt{f}\sqrt{1-v_i^2}}\, .\label{z} \ea
For a freely falling surface we get \ba\label{PPhi}
\Phi_{\sigma}(l,R)= \frac{\sqrt{1-2M/R_0}}{1-2M/R} \ea
\[
\times
\left[1 + \sigma
    \sqrt{\frac{(2M/R)(1-R/R_0)}{1-2M/R_0}}Z(l,R) ~
\right].
\]

\subsection{Bending angle}

We use the freedom in the choice of spherical coordinates to put
the angle $\theta$ in the direction to an  observer at infinity
to be equal to zero, $\theta^{(o)}=0$. Consider a null ray emitted
by the collapsing surface when its radius is $R$ and which reaches
the distant observer. Suppose its impact parameter is $l$. Then
such a ray is emitted by the collapsing surface from the region at
the angle $\theta^{(e)}$. For forward emission this bending angle
is
     \ba
    \theta_{+}^{(e)} = \Theta(l,R)\, ,
\ea \ba  \Theta(l,R)=l\int^{\infty}_{R} \frac{dr}{r^2 Z(l,r) }
\label{Theta}\, .
    \ea
For a  backward-emission  a photon before it
reaches the infinity  should pass through a turning point, $r_t < R$,
which is determined by $Z(l,r_t)=0$, or
    \ba
    r_t^2/(1-2M/r_t) &=&  l^2.  \label{turn}
    \ea
One can see that, for $l=l_{max}=  R^2/(1-2M/R)$, $r_t = R$ as
expected. Then  we get \ba
    \theta_{-}^{(e)} =2\Theta(l,r_t)-\Theta(l,R)\, .\label{btheta}
\ea

\subsection{Arrival time}

Consider a null ray with the impact parameter $l$ emitted from the
collapsing surface at the moment $\tau$ when it has the radius
$R(\tau)$. Denote by $t_{\pm}^{(o)}$  the time when it reaches a
distant observer at radius $r^{(o)}$ for  the forward/backward
ray. One has \ba
    t^{(o)}_{+}(l,\tau) = t^{(e)}(\tau)
+ \int^{r_0}_{R(\tau )} \frac{
    dr}{f(r) Z(l,r)},\label{t0f}\\
\ea \ba
    t^{(o)}_{-}(l,\tau) = t^{(e)}(\tau)
    + 2 \int^{r_0}_{r_t} \frac{dr}{f(r) Z(l,r)}
     - \int^{r_0}_{R(\tau )} \frac{dr}{f(r) Z(l,r)}
.\label{t0b} \ea It is evident that $t^{(o)}\to \infty$ when
$r_0\to \infty$. For this reason it is more convenient to consider
a finite quantity, the time difference between arrival of two null
rays emitted at two different moments of proper time, $\tau$ and
$\tau_e$, respectively. For the second ray, emitted at $\tau_e$,
we put $l=0$. Such a ray goes radially. We denote this time
difference by $\Delta t(l;\tau,\tau_e)$. In the limit when
$r^{(o)}\to \infty$ this quantity remain finite. For the forward
ray it is given by the following expression
\[
    \Delta t_{+}(l;\tau,\tau_e)
     = t^{(e)}(\tau)-t^{(e)}(\tau_e)  +T(l,R(\tau))
\]
\ba
    + R(\tau_e) -R(\tau)
    + 2M \ln
    \frac{R(\tau_e)-2M}{R(\tau) - 2M}\label{dtflr},
\ea where \ba  T(l,R) \equiv \int^{\infty}_{R} \frac{dr}{f(r)}
\left[\frac{1}{Z(l,r)}-1\right]\,. \label{TT}
     \ea
Similarly for the  backward ray one has \ba  \label{dtblr}
    \Delta t_{-}(l;\tau,\tau_e)
    =  t^{(e)}(\tau)-t^{(e)}(\tau_e)
    + 2T(l,r_t) -T(l,R(\tau))
\ea
\[
+  R(\tau_e) + R(\tau) - 2 r_t + 2M \ln
    \frac{(R(\tau)-2M)(R(\tau_e)-2M)}{(r_t -2M)^2} .
\]

The integrals for $\Theta$ and $T$ (see relations (\ref{Theta})
and (\ref{TT}), respectively) can be expressed in terms of the
elliptic functions. However, for practical calculations it is very
convenient to have approximations for these quantities in terms of
simple elementary functions. In the next section, we develop high
accuracy analytic approximations, for the integrals $\Theta(l,R)$
and $T(l,R)$.

\section{Analytic Approximation}

\subsection{Approximation for bending angle}

It is convenient to use the dimensionless quantities \ba x = M/r\,
,\hspace{0.5cm}q \equiv M/R\, ,\hspace{0.5cm}\hat{l}=l/M\, . \ea
We also denote \ba \hat{Z}=\hat{Z}(\hat{l},q)= \sqrt{1 - \hat{l}^2
q^2(1-2q)}\, , \ea so that \ba Z(l,R)=\hat{Z}(\hat{l},M/R)\, . \ea

Using these notations the integral $\Theta$, (\ref{Theta}), can be
written in the form
\ba
\Theta=
\int^q_0 dx {\hat{l} \over \hat{Z}(\hat{l},x)} \, .\label{a.1}
\ea

As already was mentioned, for a given surface radius $R$ there
exists a critical value $\hat{l}_{\max}=1/(q\sqrt{1-2q})$ and the
rays emitted from this surface must have $|\hat{l}|\le
\hat{l}_{max}$. For a fixed $q$ and $\hat{l}\in [0,\hat{l}_{max}]$
the function  $\hat{Z}$ changes from 1 to 0.

In a flat metric with $f =1$, one can calculate the integral (\ref{a.1})
analytically to get
    \ba
    \Theta_{flat} = \arccos(\sqrt{1-\hat{l}^2 q^2}).
    \ea
Leahy \cite{ll} discovered that in a wide range of its arguments
the exact integral for the bending angle can be approximated by a
simple analytical expression. A simple elegant form of the
approximative expression was proposed later by Beloborodov
\cite{bel}. In the notations adopted in the present paper this
formula reads \ba 1-\cos\beta =(1-\cos \theta^{(e)})(1-2M/R) \, ,
\ea where $\beta$ is defined by (\ref{beta}). It is easy to show
that this relation corresponds to the following approximation for
$\Theta$ \ba \Theta_0 = \arccos\left[ {\hat{Z}-2q\over 1-2q}
\right]\, \label{a.2}\, . \ea We shall refer to this relation as
to Beloborodov--Leahy (or BL--) approximation.

The typical accuracy of the BL--approximation is of order of 1\% for
the light rays emitted from the surface $R=6M$. For this reason it
works well in the study of light emitted from a surface of a neutron
star \cite{bel}. In a recent paper \cite{clm} BL-approximation was
used for fitting the light curves of X-ray pulsars. For smaller $R$
the accuracy of the BL--approximation is worse. For example it
becomes of order of 10\% for $R=4M$. In order to use this
approximation for our purposes we first slightly modify it to improve
the accuracy.

To improve the BL-approximation, let us compare expansions of
$\Theta$ and $\Theta_0$ at small values of $\hat{l}$. We have
\[
\Theta=
q\hat{l} +{1\over 12} (2-3q) q^3\, \hat{l}^3
\]
\ba + {1\over 280} (21-70q+60q^2) q^5\, \hat{l}^5+\ldots \, , \ea
\[
\Theta_0=
q\hat{l} +{1\over 12} (2-3q) q^3\, \hat{l}^3
\]
\ba +{1\over 480}(36-120q+105q^2) q^5\,\hat{l}^5+\ldots \, . \ea
($\ldots$ denotes terms of higher order in $\hat{l}$.) This gives
\ba \Theta-\Theta_0 = -{1\over 224}q^7 \hat{l}^5+\ldots \, . \ea

We write our ansatz as follows \ba\label{apan}
\hat{\Theta}(\hat{l},q)=\arccos\left[ {\hat{Z}(\hat{L},q)-2q\over
1-2q} \right]+b_5 q^2 {\cal Z}^5\, . \ea where \ba {\cal Z} \equiv
{1-\hat{Z}(\hat{l},q)\over 1-2q}. \label{calz} \ea For small
$\hat{l}$ \ba\label{ZZ} {\cal Z}= {1\over 2}q^2\,
\hat{l}^2+{1\over 16}(2-3q)\, q^4 \hat{l}^4+\ldots \, . \ea

\begin{figure}[tp]
\begin{center}
\includegraphics[height=3in,width=3.5in]{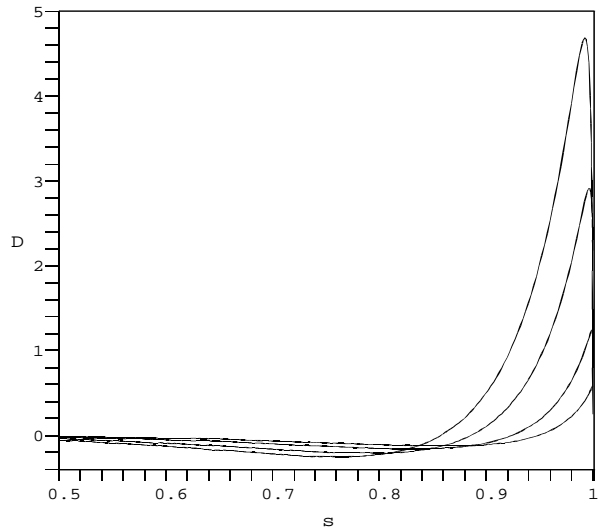}
\caption{The figure shows $D=10^3 \Delta_{\theta}$
($\Delta_{\theta}$ is a relative error for $\Theta$)  as a
function of $s=l/l_{max}$. The plots are shown for the
following 4 values of  $R/M=4.5, 5, 6, 7$ (the smaller value
$R/M$ the higher is a plot in the vicinity of $s=1$). The curves
are practically indistinguishable from $D=0$ line for $s<0.5$. For
this reason we plotted only $s\ge 0.5$ parts of the graphs. }
\label{ffig1}
\end{center}
\end{figure}

We write $b_5$ as follows \ba \label{bbb} b_5=-{\beta 2^{5/2}\over
224}\, . \ea For $\beta=1$ this expression correctly reproduces
the expansion of $\Theta(l,q)$ for small $q$ up to the terms of
the order of $O(q^7)$. Numerical calculations show that the
accuracy of the approximation is very good for the following
choice \ba \label{bbeta} \beta=3.5 \, , \hspace{0.3cm}
b_5=0.0884\, . \ea Thus the approximate expression for the
forward-emitted rays is \ba \theta^{(e)}_f\approx \hat{\Theta}(l,
q=M/R)\, , \ea where $\hat{\Theta}$ is given by (\ref{apan}),
(\ref{bbb}), and (\ref{bbeta}). The relative error
$\Delta_{\theta}=(\theta_f^{(e)}-\hat{\Theta})/\theta^{(e)}$ of
the approximate expression is very small. It is less than 0.5\%
for $R\ge 4.5 M$ for all the allowed values of $l$.
Figure~\ref{ffig1} shows the quantity $D=10^3\times
\Delta_{\theta}$ as a function of $s=l/l_{max}(q)$, where $q=M/R$.
The parameter $s$ is chosen so that for every value $q$ and the
impact parameter $l$ from $0$ to its maximum allowed value
$l_{max}$ the value of $s$ belongs to the same interval $[0,1]$.
For $R=4M$ the error $\Delta_{\theta}$ is slightly larger. It is
still less than 0.8 \%  every where excluding a narrow vicinity of
$s=1$ where it reaches 2\%.

We shall use the  formula (\ref{apan}) to approximate the bending
angle for the forward emission. For the backward emission the
approximate formula is \ba \theta^{(e)}_b(l,R)\approx  2
\hat{\Theta}(\hat{l},M/r_t)-\hat{\Theta}(\hat{l},M/R)\, , \ea
where $r_t$ is defined by (\ref{turn}).

\subsection{Approximation for arrival time}

Now we consider the arrival time. Using the dimensionless version
of $T$, ${\cal T}=T/M$, we can rewrite the expression (\ref{TT})
in the following form: \ba {\cal T}={\cal T}(\hat{l},q)= \int^q_0
dx {\hat{l}^2 \over \hat{Z}(\hat{l},x) (1+\hat{Z}(\hat{l},x))}\,
\label{t.5}. \ea We want to obtain an analytic approximation for
$T$.

Let us first assume that the function $f$ which enters $Z$ changes
slowly, and put $f=$const in the integral (\ref{t.5}). This
integral can be calculated exactly \ba {\cal
T}_0={1-\sqrt{1-\hat{l}^2\, q^2\, f}\over f \, q}\, . \ea We
restore the dependence $f$ on $q$ and use this expression with
$f=1-2q$ as a starting point for our approximation. The
corresponding expression can be written as \ba {\cal T}_0={{\cal
Z}\over q}\, . \ea One can check that this approximation is very
good for small $q$.

\begin{figure}[h]
\begin{center}
\includegraphics[height=3.0in,width=3.5in]{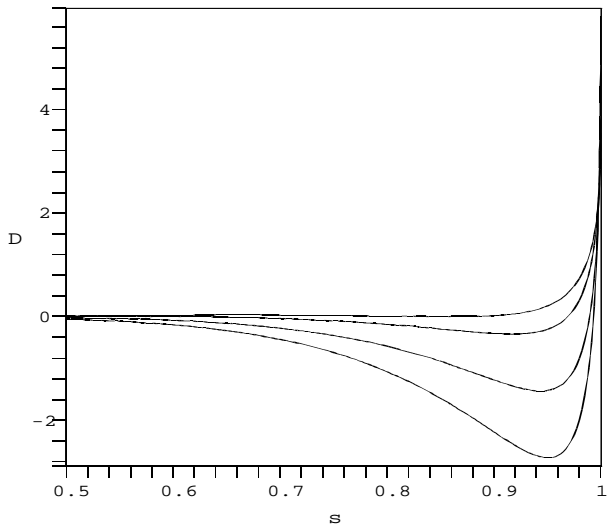}
\caption{The figure shows $D=10^3 \Delta_T$ ($\Delta_T$ is a
relative error for $T$)   as a function of $s=l/l_{max}$. The
plots are shown for the following 4 values of $R/M=4.5, 5, 6,
7$ (the smaller value $R/M$ the lower is a plot in the vicinity of
$s=1$). The curves are practically indistinguishable from $D=0$
line for $s<0.5$. For this reason we plotted only $s\ge 0.5$ parts
of the graphs.}
\label{ffig2}
\end{center}
\end{figure}

To improve it, let us compare expansions of ${\cal T}$ and ${\cal
T}_0$ at small values of $\hat{l}$. We have \ba {\cal T} = {1\over
2}q\, \hat{l}^2+{1\over 8}(q^3-2q^4)\, \hat{l}^4+\ldots \, , \ea
\ba {\cal T}_0 = {1\over 2}q\, \hat{l}^2+{1\over 16}(2q^3-3q^4)\,
\hat{l}^4+\ldots \, . \ea This gives \ba {\cal T}-{\cal T}_0 =
{1\over 16}q^4 \hat{l}^4+\ldots \, . \ea

Our ansatz for the improved approximation is \ba \hat{\cal
T}={\cal T}_0+Q({\cal Z})\, . \ea Using (\ref{ZZ}) one can
conclude that in order to provide the correct $q^4 \hat{l}^4$
behavior of the correction at small $\hat{l}$ the function $Q$
must have the following expansion \ba Q({\cal Z})= {1\over 4}
{\cal Z}^2+\ldots \, . \ea We write our ansatz as follows \ba
Q({\cal Z})={1\over 4} {\cal Z}^2+a_3 {\cal Z}^3+a_4 {\cal Z}^4\,
. \label{hatt} \ea Numerical calculations show that the accuracy
of the approximation is very good for the following choice of the
parameters \ba a_3=1/15\, ,\hspace{0.5cm}a_4=1/25\, . \ea Thus the
approximation for the time delay takes the form \ba\label{APPROX}
\hat{\cal T}={ {\cal Z}\over q}+Q({\cal Z})\, ,\hspace{0.3cm}
Q({\cal Z})={1\over 4} {\cal Z}^2+{1\over 15} {\cal Z}^3+{1\over
25} {\cal Z}^4\, , \ea \ba \hat{T}(l,R)=M \hat{\cal
T}(\hat{l},M/R)\, . \ea

The relative error $\Delta_T=(T-\hat{T})/{T}$ of the approximate
expression is again very small. It is less than 0.5\% for $R\ge
4.5 M$ for all the allowed values of $l$. Figure~\ref{ffig2} shows
the quantity $D=10^3\times \Delta_T$ as a function of
$s=l/l_{max}$. For $R=4M$ the error $\Delta_T$ is slightly larger.
It is still less than 1 \% every where excluding a narrow vicinity
of $s=1$ where it reaches 3\%.

For the calculations of the intensity of the light from a
radiating body observed at infinity we shall need the expression
for the derivatives $\hat{\cal T}_{,q}$ and $\hat{\cal
T}_{,\hat{l}}$. The first one can be easily found from the
definition of ${\cal T}$ (\ref{t.5}) \ba \hat{\cal
T}_{,q}={\hat{l}^2 \over \hat{Z}(\hat{l},x)
(1+\hat{Z}(\hat{l},x))}\, . \label{Tl} \ea

\begin{figure}[h]
\begin{center}
\includegraphics[height=3.0in,width=3.5in]{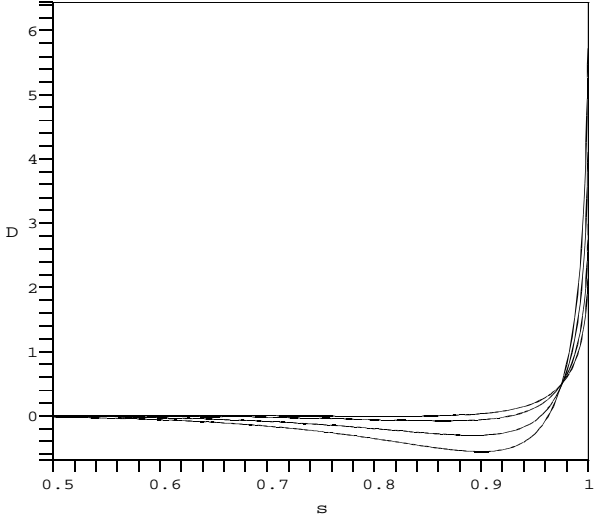}
\caption{The figure shows $D=10^2 \Delta_{T_{,l}}$ ($\Delta_{T_{,l}}$
is a relative error for $T_{,l}$)   as a function of $s=l/l_{max}$.
The plots are shown for the following 4 values of $R/M=4.5, 5, 6,
7$. In the region where $D<0$, the smaller value $R/M$ the lower is a
plot in the vicinity of $s=1$. The curves are practically
indistinguishable from $D=0$ line for $s<0.5$. For this reason we
plotted only $s\ge 0.5$ parts of the graphs.}
\label{ffig2_l}
\end{center}

\end{figure}

For the derivative $\hat{\cal T}_{,\hat{l}}$ we shall use the
approximate expression obtained by the differentiation of
$\hat{\cal T}$ given by (\ref{APPROX}) with respect to $\hat{l}$.
The approximation for the derivative $\hat{\cal T}_{,\hat{l}}$
works slightly worse than the approximation for $\hat{\cal T}$. We
denote \ba \Delta_{T_{,l}}=({\cal T}_{,\hat{l}}-\hat{\cal
T}_{,\hat{l}})/ {\cal T}_{,\hat{l}} \ea the relative error. The
figure~\ref{ffig2_l} shows $D=10^2 \Delta_{T_{,l}}$ as a function
of $s=l/l_{max}$. The maximum value of the error (for $R=4.5 M$)
is near $s=1$ and it reached 5\%.

\section{A short flash from a collapsing surface}

\subsection{Flux and intensity for a short flash}

We shall use superscripts $(e)$ and $(o)$ for  emitted and
observed radiation, respectively. To characterize such a radiation
one uses the notions of the {\em flux} and {\em intensity}. The
flux, $F$, is  a power of the radiation  per unit area emitted by
a radiating surface or received by the detector. The intensity,
$I$, is a power of either emitted or observed radiation per unit
area per unit solid angle. Quite often the spectral
characteristics, called the {\em specific flux}, ${\cal F}_{\nu}$,
and {\em specific intensity}, ${\cal I}_{\nu}$, are also used.
They are "per unit frequency" versions of $F$ and $I$,
respectively. Evidently, one has \ba\label{specif} F=\int d\nu
{\cal F}_{\nu}\, ,\hspace{0.3cm} I=\int d\nu {\cal I}_{\nu}\, .
\ea The intensity $I$ and specific intensity ${\cal I}_{\nu}$
depend on the angle $\theta$ between the direction of radiation
and the normal to emitting (observing) area. The flux and specific
flux are obtained by integration of these quantities over the
directions \ba\label{FF} F = \int  d\Omega I\, ,\hspace{0.3cm}
{\cal F}_{\nu} = \int  d\Omega {\cal I}_{\nu}\, , \ea where
$d\Omega$ is an element of a solid angle. For an isotropic case
with azimuthal symmetry, \ba d\Omega=2\pi \sin\theta d\theta \, .
\ea

The observed intensity $I^{(o)}$ depends on the direction, or (in
our case) on the angle $\theta$ between the normal to the detector
surface and the direction of observation. Instead of the direction
angle $\theta$ one can use the impact parameter $l$  of an
incoming photon. For large $r_0$ one has $\theta=l/r_0$, and
$d\Omega^{(o)}=2\pi l dl/r^2_0$. Thus the relation (\ref{FF}) can
be written as \cite{remark} \ba
      F^{(o)} = \int l dl {2\pi I^{(o)}\over r_0^2}\, \label{F0}.
\ea
In the integrand,  the
intensity $I^{(o)}$  is now considered as a
function of impact parameter $l$:  $I^{(o)}=I^{(o)}(l)$.

The impact parameter $l$ is conserved along the photon trajectory.
It specifies the trajectory itself. Thus for a given value $l$ one
can determine both, the bending angle $\theta^{(e)}$ (the angle on
the surface of a collapsing body where the ray was emitted) and
the angle $\beta$ at which it was emitted.

One can show (see e.g. Exercise 22.17 in \cite{mtw}), that the
quantity $I_{\nu}(\nu)/\nu^3$ remains constant along a photon's
world line. This quantity is proportional to the number density in
phase space for photons and its conservation follows from the
Liouville's theorem (see e.g., \cite{at} and \cite{pod}).

Consider a light ray with the impact parameter $l$ emitted from
the collapsing surface at the moment of the proper time  $\tau$,
and let $t$ be the time when it reaches a distant observer at
$r_0$. For fixed $r_0$ relations (\ref{t0f}) and (\ref{t0b}) can
be used to determine $\tau=\tau(t,l)$.

The specific flux as measured by a distant observer at time  $t$
is \ba {\cal F}^{(o)}_{\nu_0}(t)= {2\pi \over r_0^2} \int l dl
\Phi^{-3}  {\cal I}_{\nu_e}^{(e)}(l,\nu_e,\tau(t,l))\,
.\label{sflux} \ea Here $\Phi=\Phi(l,R)$ is given by eq.(\ref{z}),
and $R$ is the radius of the collapsing surface at the moment when
a null ray with the impact parameter $l$ leaves it in order to
reach the observer at $r_0$ at the moment $t$. We denote by $\tau$
the corresponding proper time on the surface. For a fixed time of
arrival, $t$, and the moment of emission $\tau$, one can use the
relations (\ref{t0f}) and (\ref{t0b}) to determine the
corresponding value of the impact parameter $l=l_t(\tau)$. Thus
the integral over $l$ in eq.(\ref{sflux}) can be rewritten  as an
integral over the proper time,   $\tau$: \ba  {\cal
F}^{(o)}_{\nu_0}(t)= \frac{2\pi}{r_0^2} \int  d\tau W \Phi^{-3}
{\cal I}_{\nu_e}^{(e)}(l,\nu_e,\tau) \, , \label{fnu} \ea where
\ba W\equiv l \left|dl\over d\tau \right|\, . \ea

The expression for $W$ can be obtained by differentiating
eq.(\ref{t0f}) and eq.(\ref{t0b}) with respect to $\tau$ for a
fixed value of $t^{(o)}_{\sigma}$.  Using the analytic
approximations for the arrival time, we get (for details see
Appendix) \ba\label{W1} W={\cal J}^{-1}_{\sigma}\,| Z- \sigma v_i|
R {dt^{(e)}\over d\tau}\, . \ea Here $dt^{(e)}/d\tau$ is given by
(\ref{tetau}), and \ba\label{W2} {\cal J}_{+} =   1 +
   \frac{M}{R}Q'({\cal Z})\, ,
\ea \ba\label{W3} {\cal J}_{-} =   {\cal J}_{+}
   + {4M R Z\over r_t^2}
  \frac{(r_t+M Q'_{r_t})}{r_t - 3M}\, .
\ea Here $Q'_{r_t}=Q'(f^{-1}(r_t))$, $Z=Z(l,R)$, ${\cal Z}={\cal
Z}(l,R)$, and  $\Phi=\Phi(l,R)$, and the arguments in these
functions are  $l=l_t(\tau)$ and $R=R(\tau)$. We also use a
notation $Q'=dQ/d{\cal Z}$. A transition from the forward to the
backward ray regime occurs at the point where $Z=0$ that is at
$l=l_{max})$. It is easy to see that at this point $W_-=W_+$, as
it is expected.

We shall need an explicit form of $W$ for a special case when a
ray is emitted radially and hence $l=0$. In this case only forward
emission is possible and we have ${\cal J}_{+} = 1$. Thus the
corresponding value $W_0$ takes the form \ba\label{WWW0}
W_0=\left[1+\sqrt{\frac{2M}{R}}
     \frac{\sqrt{1-R/R_0}}{\sqrt{1-2M/R_0}}\right]\,
     {R\sqrt{1-2M/R}\over 1-2M/R_0} \, .
\ea It is also easy to check that for a tangentially emitted ray
($l=l_T$) $W$ vanishes.

For  a very short  in time flash from the surface at the moment
$\tau_e$, the intensity can be  approximated  as \ba {\cal
I}_{\nu_e}^{(e)}(l,\nu_e,\tau) = I_{\nu_e}^{(e)}(l,\nu_e)
\delta(\tau-\tau_e). \label{ftau} \ea In this case the integration
over $\tau$ in (\ref{fnu}) is trivial and we get for the observed
specific  flux ${\cal F}_{\nu_0}^{(o)}$ the following expression
   \ba\label{FFFF}
     {\cal F}_{\nu_0}^{(o)}(t) &=& {2\pi\over r_0^2} W_e \Phi_e^{-3}
    I_{\nu_e}^{(e)}(l_e,\nu_e)\, .
   \ea
Here $ R_e = R(\tau_e)$, $l_e=l_t(\tau_e)$, $W_e=W(l_e,R_e)$, and
$\Phi_e=\Phi(l_e,R_e)$. The factors $W_e$ and $\Phi_e$, which
enter this relation, do not depend on the frequency. Thus for the
observed flux we have \ba F^{(o)}(t)=\int d\nu_0 {\cal
F}_{\nu_0}^{(o)}(t)=\int d\nu_e \Phi_e^{-1} {\cal
F}_{\nu_0}^{(o)}(t)\, . \ea Using relations (\ref{specif}) and
(\ref{FFFF}) one obtains \ba F^{(o)}(t)={2\pi\over r_0^2} W_e
\Phi_e^{-4}
    I^{(e)}(l_e)\, , \label{fot}
\ea where the intensity $I^{(e)}(l)$ is \ba I^{(e)}(l)=\int d\nu_e
I_{\nu_e}^{(e)}(l_e,\nu_e)\, . \ea

It should be emphasized that the simplicity of the expression
(\ref{fot}) is a consequence of the assumption that the pulse has
a $\delta$-like form. In a general case the integration over the
time and frequencies cannot be performed explicitly. For this
reason, a so-called monochromatic approximation is often used in
the discussions of the continuous radiation from relativistic
objects (see e.g. \cite{bel}). We note also that in the adopted
approximation all the expressions in (\ref{fot}) are defined in an
explicit form in terms of elementary functions. This radically
simplifies the study of the light curves.

\subsection{Normalized flux}

We denote the flux registered by the distant observer at the
moment $t$ as $F^{(o)}(t)$ and  we denote by  $F^{(o)}(0)$ the
flux  at the moment when the first ray arrives to the distant
observer. Such a ray is emitted forwardly ($\sigma=+$ and $l=0$)
at $\theta_e=0$. It is convenient to normalize the observed
time-dependent flux $F^{(o)}(t)$ to the value $F^{(o)}(0)$. We
denote this ratio \ba {\bf F}(t)={F^{(o)}(t)\over F^{(o)}(0)}\, .
\ea Using the relation (\ref{fot}) one obtains \ba \label{FBF}
{\bf F}(t)={W_e(t)\over W_0} \left({\Phi_e(t)\over
\Phi_0}\right)^{-4} {\bf I}\, , \ea \ba {\bf
I}=\frac{I^{(e)}(l_e)}{I^{(e)}(0)}\, . \ea The explicit
expressions for the `first-ray' quantity $W_0$ is given by
(\ref{WWW0}). The equation (\ref{PPhi}) determines $\Phi_e$ and
$\Phi_0$. In the latter case one needs to put $\sigma=+$ and
$Z=1$. Relations (\ref{W1}), (\ref{W2}), and (\ref{W3}) allow one
to find $W_e$.

Let us discuss now the last factor which enters the expression for
${\bf F}(t)$. In general, the intensity of the radiation from the
surface of star can be written \cite{flux} as \ba I^{(e)}(l)= a +
b \cos(\beta(l)), \ea where $a$ and $b$ depend on the details of
the emission process. In this work,  we calculate two extreme
cases: (A)  $\, a \neq 0,\, b=0$,  and (B)  $\, a = 0,\, b \neq
0$. The former corresponds to the isotropic emission from a
optically thick object and the latter corresponds to Lambert's
law. For $l=0$ one has $\beta(l)=0$ and $\tilde{I}^{(e)}= a + b$,
so that \ba {\bf I}={a + b \cos(\beta(l))\over a + b}\, . \ea In
the first case ${\bf I}_A=1$, while in the second one ${\bf
I}_B=\cos(\beta(l))$.

\section{Redshift and light curves for a flash from a collapsing surface}

To illustrate the obtained results, we consider now special
examples. As a first example, we consider a neutron star which
looses its stability. In this case an initial radius $R_0$  is
$R_{NS}  = 12 - 20$~ km and the mass is of order of $M \sim 1.5
M_{\odot}$ \cite{lp2}, and hence $ R_0/M = 5.4 - 9 $. Another
example is a proto neutron star $R_{PNS} \sim 20$~km and $M \sim
1.5 M_{\odot}$ \cite{lp1}. In this case $ R_0/M = 9 $. In this
section, we discuss in detail two cases with initial radii: $R_0
=5.4M$ and $R_0 =9 M$.

For a freely falling surface with $R_0 =5.4 M$, the turning point
$r_t$ on the trajectory  of a backward ray lies within the valid
range of  the analytic approximation, $ r_t >4.5M$, provided $R_e \ge
4.8M $. In accordance with this we choose $R_e=4.8 M$ (case I).

For a freely falling surface with $R_0 =9M$, the analytic
approximation can be applied to the emission at $R_e \ge 5.5M$. In
this case we calculate a bending angle, redshift and  a fluxes
registered by a  distant observer for the following 3 values of
$R_e/M = 5.5, 6.5, 7.5$ (cases IIa, IIb, and IIc, respectively).

\subsection{Arrival time}

Since the arrival time depends on the position of the radiating
region on the surface, a light emitted at the same moment $\tau_e$
from the surface $R_e$ reaches a distant observer through some
finite interval of time $t$. During this period, the observed
brightness and frequency are changing. First ray arriving to the
distant observer is emitted from $\theta_e=0$ and it has $l=0$ and
$\sigma=+$. The time when this first ray reaches a distant
observer depends on the position $r_0$ of this observer. This time
becomes infinite when $r_0\to \infty$. It can be made finite if,
for example, instead of $t$ one uses the retarded time $u=t-r^*$,
where $r^*$ is the tortoise coordinate. We choose $u=0$ for the
moment of arrival of the first ray.  At given radius $\Delta
u=\Delta t$. The maximum arrival time difference  is  assigned for
the backward ray emitted  with an impact parameter $l_T$, and it is
\ba
    \Delta t_{max} &=&  \Delta t_{-}(l_T;\tau_e,\tau_e)
    =
     2T(l_T,r_t) -T(l_T,R_e), \nonumber \\
     & & +  2 R_e  - 2 r_t + 4M \ln
    \frac{(R_e-2M)}{(r_t -2M)} . \label{dtmax}
\ea

In the case I, for $R_0/M=5.4 $ and $R_e/M = 4.8$, the time delay
is calculated is $\Delta t_{max}/M=13.8$.  In the case II,
for $R_0/M = 9$ the time delay for different values of $R_e$ is given
in the Table.

\medskip

\begin{table}[h]{\bf Time delay for $R_0/M = 9$}\\
\begin{tabular}{|c|c|c|c|}\hline
Case & IIa & IIb & IIc \\
\hline\hline
$R_e/M$ & 5.5 & 6.5 & 7.5\\
\hline
$\Delta t_{max}/M$ & 16.9 & 15.4 & 14.4\\
\hline
\end{tabular}
\end{table}

In what follows it is convenient to  use  a normalized arrival time
difference defined as $\delta  \equiv \Delta t / \Delta t_{max}$. We
shall call this quantity the time parameter. The time parameter is
always changes in the interval $[0,1]$. The  time parameter for
forwardly emitted light   increases as $l$ increases from $l=0$ to
$l_{max}$. The backward emission starts with $l_{max}$ and ends at
$l_T$ and the time parameter for a backward emission is increasing as
$l$ changes from $l_{max}$ to $l_T$.

\subsection{Bending angle and redshift}

\begin{figure}[htp]
\begin{center}
\includegraphics[height=2.5in,width=2.5in]{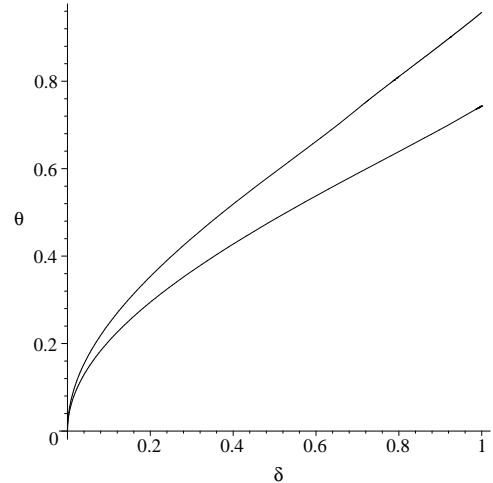}
\caption{Bending angle as a function of the time parameter $\delta$.
The upper curve shows the bending angle  for a freely-collapsing
surface for $R_0 =5.4M$ and $R_e = 4.8M$. The lower curve show the
bending angle  for  a static surface with $R_e = 4.8M$.}
\label{bangle1}
\end{center}
\end{figure}

The bending angle as a function of the time parameter is a
monotonously increasing function. For the case I it is shown in
Fig.~\ref{bangle1} (an upper curve). For a comparison  a similar
function for a static surface (of the radius $R_e = 4.8M$) is shown
at the same Figure (a lower curve).

\begin{figure}[htp]
\begin{center}
\includegraphics[height=2.5in,width=2.5in]{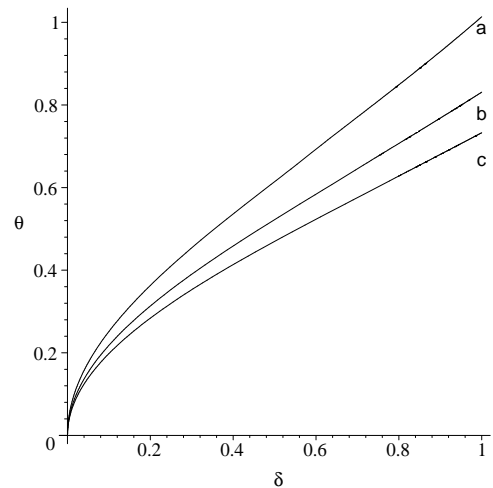}
\caption{Bending  angle  as a function of the time parameter $\delta $.
The three graphs show the bending angle for the same initial
radius $R_0 =9M$  for 3 different values of $R_e$:  (a)
$5.5M$, (b)  $6.5M$, and (c) $7.5M$.}
\label{bangle2}
\end{center}
\end{figure}

Figure~\ref{bangle2} demonstrates the bending angle as a function of
the time parameter for 3 different cases IIa, IIb, and IIc. The
smaller is $R_e/M$, the faster is the radial motion of the radiating
surface, and the larger is the observed region with the backward
emission. As a result the range of bending angle for smaller values of
$R_e/M$ becomes larger, as it can be seen at the Figure~\ref{bangle2}.

\begin{figure}[tp]
\begin{center}
\includegraphics[height=2.5in,width=2.5in]{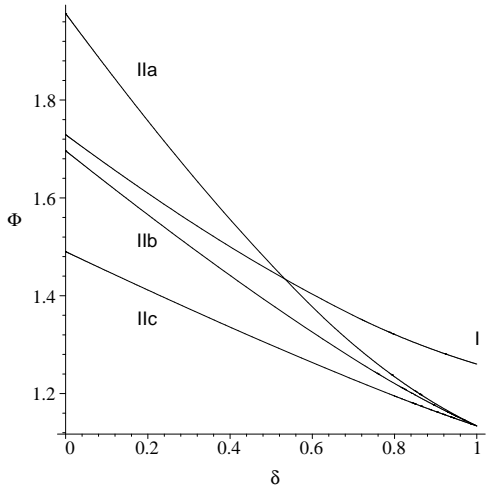}
\caption{Redshift factor  for  a freely collapsing  surface as a
function of the time parameter $\delta $ for the cases I, and
IIa,b,c.} \label{shift1}
\end{center}
\end{figure}

The frequency observed at  infinity is different from the frequency
at emission  because of two reasons: (1) Difference of  the
gravitational potential at the point of emission and observation,
(gravitational redshift), and (2) The velocity of  the emitting
surface (Doppler shift). The  photons emitted from the surface of
$R_e$  experience the same gravitational redshift independent of
their  angular positions (bending angle) of emission. However Doppler
shift depends on the relative velocity of the surface of emission
with respect to the distant observer, and hence it depends  on the
bending angle (or the impact parameter $l$).   The corresponding
total frequency shift can be calculated from eq.(\ref{PPhi}). Since
the arrival time depends on the impact parameter as well, the
frequency shift then can be plotted as a function of the arrival
time. The calculated ratio of   emitted frequency to the  observed
one, $\Phi$,  for a short flash as a function of the time parameter
$\delta $ is shown in Fig.~\ref{shift1}. Three curves which meet one
another at $\delta =1$ correspond to the three cases IIa,b,c. The
forth curve corresponds to the case I.

It is interesting that the redshift due to the gravity is
substantially cancelled by the Doppler shift for the tangentially
emitted light. Really, using (\ref{lT}) and (\ref{PPhi}) we obtain
for the redshift factor $\Phi$ the following expression \ba
\Phi_T={\sqrt{1-v_i^2\over f(R_e)}}\, . \ea For a free fall from
the radius $R_0$ one has \ba 1-v_i^2={1-2M/R\over 1-2M/R_0}\, .
\ea Thus \ba \label{RST} \Phi_T = \frac{1}{\sqrt{1-2M/R_0}}\, .
\ea It means that for the ``last rays" (that is for rays with
$l=l_T$), the redshift depends only on the initial radius $R_0$
and does not depend on the radius of emission $R_e$. For this
reason the three curves IIa,b,c in Fig.~\ref{shift1} merge at the
same value $3/\sqrt{7}\approx 1.134$ at $\delta  =1$ (that is for
$l=l_T$). Relation (\ref{RST}) also shows that  for $R_0=\infty$
the gravitational redshift is exactly cancelled by the Doppler
shift \cite{jaffe}.

Since for direct radial rays ($l=0$) both effects ``work" in the
same direction, one can expect that for a given $R_0$ the redshift
will be larger for smaller values of $R_e$. The
Fig.~\ref{shift1} clearly demonstrates this.

\subsection{Light curves}

\begin{figure}[tp]
\begin{center}
\includegraphics[height=2.5in,width=2.5in]{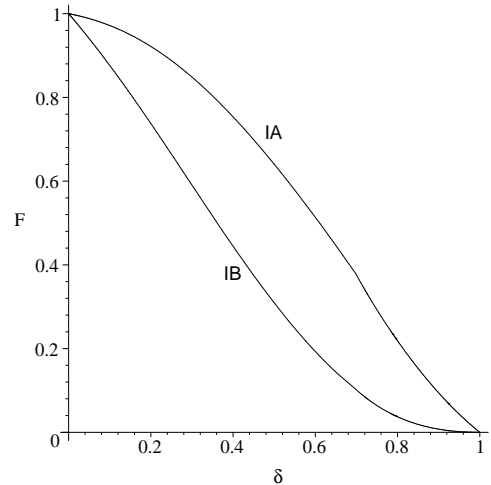}
\caption{Light curves for ${\bf I}=1$ (IA) and ${\bf I}=
\cos\beta(l)$ (IB)
for a freely-collapsing surface with $R_e/M=4.8$ and  $R_0/M
=5.4M$.}
\label{flux54ab}
\end{center}
\end{figure}

\begin{figure}[tp]
\begin{center}
\includegraphics[height=2.5in,width=2.5in]{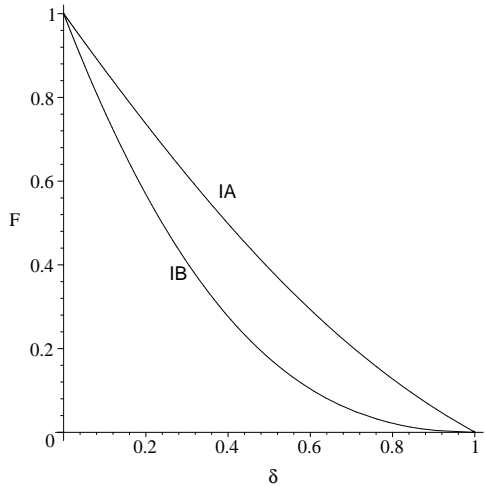}
\caption{Light curves for ${\bf I}=1$ (IA)  and ${\bf I}= \cos\beta(l)$
(IB)  for a static surface with $R_e/M=4.8$.} \label{flux5sab}
\end{center}
\end{figure}

Let us discuss now normalized flux  as a function of the time
parameter $\delta $. We call the corresponding graph a light
curve. In case I ($R_0=5M$, $R_e= 4.8M$), the  light curves for a
short flash from the collapsing spherical surface  are shown
Fig.(\ref{flux54ab}). The plot IA is a light curve for ${\bf
I}=1$ and the plot IB is a light curve for ${\bf
I}=\cos\beta(l)$ (Lambert's law). Similar light curves for a static
spherical surface are shown for comparison at Fig. \ref{flux5sab}.

\begin{figure}[htp]
\begin{center}
\includegraphics[height=2.5in,width=2.5in]{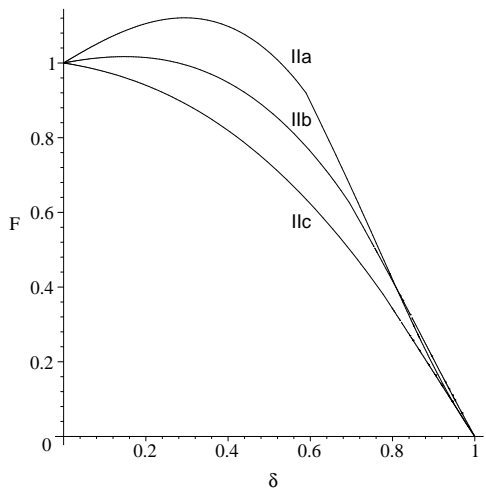}
\caption{Light curves for ${\bf I}=1$ for the cases IIa,b,c.}
\label{flux9a}
\end{center}
\end{figure}

\begin{figure}[htp]
\begin{center}
\includegraphics[height=2.5in,width=2.5in]{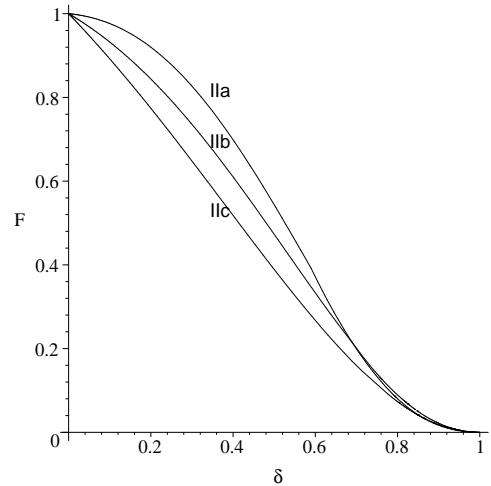}
\caption{Light curves for ${\bf I}=\cos\beta(l)$ for the cases IIa,b,c.}
\label{flux9b}
\end{center}
\end{figure}

For the case II ($R_0 = 9M$) the light curves for A and B type of the
radiating surface are shown in Fig.~\ref{flux9a} and
Fig.~\ref{flux9b}, respectively. Each of the figures contains 3
curves corresponding to IIa,b,c cases.

Let us discuss now qualitative behavior of the light curves. The
observed normalized flux ${\bf F}(t)$ given by eq. (\ref{FBF}) is a
product of 3 factors: (1) a kinematic term $W_e(t)/W_0$, (2) a
redshift factor $(\Phi_e(t)/\Phi_0)^{-4}$,  and (3) a normalized
intensity of the emission ${\bf I}$. The third factor depends on the
model of the radiating surface and it does not depend on the arrival
time. The first two factors are time dependent. The arrival time
dependence of $W_e$ is essentially determined by the factor of $|Z-
\sigma v_i| $ in eq.(\ref{W1}), which is a decreasing function of
$\delta $ and vanishes  for $\delta  = 1$.  Hence every light curves
should cross the zero-flux axis  at $\delta =1$.  For  a static
surface $v_i =0$ and $Z$ (and hence $W_e(t)$) vanishes at $\delta =1$,
where $l=l_{max}$.  For a collapsing surface  $v_i <0$ and the
observable flux vanishes not for $l_{max}$ (where $Z=0$) but for the
backward emission with  $l=l_T$.  Hence one can expect longer
duration of observed flux for the emission from a collapsing surface
compared to the emission from a static surface.  The effect of motion
of the collapsing surface becomes stronger for larger $v_i$. For
example for a given $R_0=9M$, $\Delta t_{max}$ is calculated to be
larger for smaller  $R_e$ for which $v_i$ is larger (see Table).

The redshift factor $\Phi$  depends basically on the relative
receding velocity of the emitting region (determined by the
bending angle) with respect to the distant observer.  The relative
receding velocity  is decreasing  as  the bending angle is
increasing. Since the arrival time difference  becomes larger for
a ray with  larger bending angle,  one can expect the enhancement
of a factor, $\Phi^{-4}$, for larger $\delta $.   The main effect
of the frequency shift for the observed flux due to the collapsing
surface  is the enhancement of the flux for lately arriving rays.
As a result  the  shape of the light curve for a collapsing
surface is changing from that of a static surface in such way that
the  flux decreasing in  $\delta $ is delayed  and the sharp
forward peak at $\delta =0$ becomes a rather  smooth peak as
shown in Fig.~(\ref{flux54ab}). For sufficiently large collapsing
velocity, for example for $R_0=9M$ and $R_e=5.5M$, one can observe
that position of the peak in the light curve also changes  from
$\delta =0$ to a later arrival time $\delta  \neq0$ for the
isotropic intensity profile(A) as shown at Fig.~(\ref{flux9a}).
The emission angle with respect to the normal to the surface,
$\beta$, varies  from 0 to $\pi/2$ as the impact parameter $l$
varies from 0 to $l_{max}$ and further to $l_T$. Hence the
intensity profile of (B) with $I^{(e)}=b\cos(\beta(l))$ suppresses
the enhancement due to the factor  $\Phi^{-4}$ for  lately
arriving rays substantially as shown in Fig.~\ref{flux9b}.  The
light curves from the static surface for intensity profile (A) and
(B) are also shown in Fig.~\ref{flux5sab} for comparison.

\section{Discussion}

In this work, we  discussed characteristics of the radiation emitted
from a  surface of a collapsing object. We  studied a simplified
(toy) model in which a radiation of massless particles has a sharp in
time pulse profile and it happens at the surface at the same instant
of time (from a point of view of a comoving observer). In this
approximation both integrals over the time and frequency which enter
a general expression for the observed flux can be taken. As a result
we obtained an expression for the normalized flux as a function of
the time of the observation. We demonstrated that for a  short in
time flash, the observed normalized flux can be expressed as a
product of  three terms: a kinetic term ($W$), a redshift factor
($\Phi$), and the  intensity of the emitted radiation. The intensity
of the emitted radiation is model dependent. In particular it depends
on the model of the radiating surface region. The other two factors
are universal and depend only on the equation of motion of the
collapsing surface. The dependence of $W$ and $\Phi$ on the arrival
time is quite different. We assume that the collapse starts at some
radius $R_0$ and the emission occurs at the radius $R_e$. Under this
assumption all the characteristics of the radiation such as its
duration, redshift, and the form of the light curves (for chosen
intensity of emission) depend only on these 2 parameters. To obtain
the light curves one needs to integrate the equations for the light
propagation in the Schwarzschild metric. The corresponding
integration can be performed in terms of elliptic integrals. But
the calculation of the flux requires solving an inexplicit
equation which contains the elliptic integrals.

In order to solve this problem  we developed an analytical
approximation for the bending angle and time delay for null rays
emitted by a collapsing surface. In the case of the bending angle
this approximation is an improved version of the earlier proposed
BL-approximation \cite{ll,bel}. For rays emitted at $R > 2R_g$ the
accuracy of the approximation for the bending angle and time delay
proposed in the present paper is of order (or less) than 2-3$\%$. By
using this approximation we obtained an explicit formula for the
observed normalized flux, which not only allows one very efficient
numerical calculations of the characteristics of the radiation, but
also appears to be useful for understanding the qualitative features
of the radiation.

Even for a static object the effects of the General Relativity allows
one to "see" a part of its opposite side surface. For a collapsing
object this effect is more profound. As a result,  the duration of
the flux is elongated. Another difference is that the sharp decrease
in time for the static surface is delayed and "smoothed out" so that
the peak becomes broader.  For a sufficiently large collapsing
velocity the peak position can even be shifted to $\delta  >0$. We
demonstrated these features by considering two examples of
collapses starting at $R_0=5.4M$ and $R_0=9M$.

Though  in this paper we focused on a model of brief in time
flash emission, some of its results (improved analytic
approximation) might be of the interest for other astrophysically
interesting problems.

\noindent
\section*{Acknowledgment}\noindent
This work was supported by the Natural Sciences and Engineering
Research Council of Canada and by the Killam Trust. HLK was
supported also in part by Korea Research Foundation
Grant(KRF-2004-041-C00085).  HKL would like to thank Valeri Frolov
for the kind hospitality during his visit to University of Alberta
and also Roger Blandford for his kind invitation  to KIPAC and
thanks Andrei Zelnikov for discussions.

\appendix

\section{Calculation of ${dl}/{d \tau}$}

Let us fix $\tau_e$ in (\ref{dtflr}) and consider time delay
$\Delta_+(l;\tau,\tau_e)$ as a function of $l$ and $\tau$. Consider a
rays which reach the observer at the same time $t$. For these rays
$\Delta_+(l;\tau,\tau_e)$ is the same. This establish the relation
between $l$ and $\tau$: $l=l_t(\tau)$. To obtain this expression we
differentiate (\ref{dtflr}) with respect to $\tau$, keeping $t$ fixed.

For a forward ray one has \ba \dot{t}^{(e)}-{\dot{R}\over
fZ}+{\partial T\over \partial l}\dot{l}=0\, . \ea We denote by a
dot the derivative $d/d\tau$. Using (\ref{RRRR}) we get \ba
\dot{l}=-{ (Z-v_i)\dot{t}^{(e)}\over Z (\partial T/\partial l)}\,
. \ea To calculate the partial derivative \ba {\partial T\over
\partial l}={\partial T\over \partial {\cal Z}} {\partial {\cal
Z}\over \partial l} \ea we use the approximate expression
(\ref{APPROX}). We have \ba {\partial T\over \partial {\cal
Z}}=R+M Q'\, , \ea \ba {\partial {\cal Z}\over \partial l}={l\over
R^2 Z}\, . \ea Combining these results we obtain \ba
W_+=l|\dot{l}|={R(Z-v_i)\dot{t}^{(e)}\over {\cal J}_+}\, . \ea
where \ba {\cal J}_+= 1+(M/R) Q'\, . \ea

For backward rays the calculation of $\dot{l}$ is similar. Using
the equation (\ref{dtblr}) for rays reaching the observer at the
same time $t$ one obtains the following equation for $l_t(\tau)$
\ba\label{eqb} \dot{t}^{(e)}+{\dot{R}\over f(R)Z(l,R)}-{\partial
T\over \partial l}\dot{l}+2\left[ \dot{T}(r_t)-{\dot{r}_t\over
f(r_t)}\right]=0\, . \ea Using (\ref{turn}) we find \ba
\dot{r}_t={l\dot{l} f^2(r_t)\over (r_t-3M)}\, . \ea We also have
\ba Z(r_t)=0\, ,\hspace{0.5cm} {\cal Z}(r_t)=1/f(r_t)\, . \ea
Using the approximate expression for $T$, (\ref{APPROX}), we have
\ba T_{,r_t}={\cal Z}(r_t)+ [r_t +MQ'_{r_t}] {d{\cal Z}\over
dr_t}\, . \ea Here $Q'_{r_t}=Q'(f^{-1}(r_t))$.
 Using these
relation we obtain \ba \dot{T}(r_t)-{\dot{r}_t\over
f(r_t)}=-l\dot{l} {2M\over r_t^2(r_t-3M)}[r_t+MQ'_{r_t}]\, . \ea
This allow us to rewrite (\ref{eqb}) in the form \ba
\dot{t}^{(e)}(Z+v_i)R=l\dot{l} \left[{\cal J}_+ +{4MRZ\over
r_t^2}{r_t+MQ'_{r_t}\over r_t-3M}\right]\,. \ea Finally we obtain
\ba W_-=l|\dot{l}|={R(Z+v_i)\dot{t}^{(e)}\over {\cal J}_-}\, , \ea
\ba {\cal J}_-={\cal J}_+ + {4MRZ\over r_t^2}{r_t+MQ'_{r_t}\over
r_t-3M}\,. \ea


\begin{thebibliography}{}
\bibitem{remark} In the presence of both, forward and backward rays,
the integral over the impact parameter $l$ is understood as a sum
of two integrals, one  over $l\in [0,l_{max}]$ for the forward
radiation and another over $l\in [l_T,l_{max}]$ for the backward
radiation. Thus the integral (\ref{F0}) stands for \ba F^{(o)} =
{2\pi \over r_0^2} [\int_0^{l_{max}} l dl\,  I_+^{(o)}
+\int_{l_T}^{l_{max}} l dl\,  I_-^{(o)}] \, . \ea We shall use
this convension for the other similar integrals.
\end{thebibliography}
\end{document}